\documentclass[12pt]{iopart}
\usepackage{amsmath}
\usepackage{iopams}

\begin{document}

\maketitle

\title{Bateman Oscillators: Caldirola-Kanai and Null Lagrangians and Gauge Functions}

\author{L.C. Vestal and Z. E. Musielak}
\address{Department of Physics, The University of Texas at 
Arlington, Arlington, TX 76019, USA}
\ead{lesley.vestal@mavs.uta.edu; zmusielak@uta.edu}

\begin{abstract}
The Lagrange formalism is developed for Bateman oscillators,
which include both damped and amplified systems, and a novel method 
to derive the Caldirola-Kanai and null Lagrangians is presented.  For the 
null Lagrangians, corresponding gauge functions are obtained.  It is 
shown that the gauge functions can be used to convert the undriven 
Bateman oscillators into the driven ones.  Applications of the obtained 
results to quantizatation of the Bateman oscillators are briefly discussed.
\end{abstract}


\section{Introduction} 

The Bateman model consists of two uncoupled oscillators, damped 
or time-forward and amplified or time-reversed [1], which are called 
here the Bateman oscillators.  The equations of motion for these 
oscillators are derived from one Lagrangian that is known as the 
Bateman Lagrangian [1,2] and used in studies of the damped 
harmonic oscillator and its quantization [3,4].  General solutions 
of the equations of motion for the Bateman model are well-known 
and given in terms of elementary functions [5].  

We solve the inverse problem of the calculus of variations [6] for the 
Bateman oscillators and develop a novel method to derive the standard 
and null Lagrangians.  The standard Lagrangians (SLs) contain the square 
of the first order derivative of the dependent variable (the kinetic energy 
like term) and the square of dependent variable (the potential energy like
term), and among the derived SLs, the Caldirola-Kanai (CK) Lagrangian 
is obtained [7,8] and its validity to describe the Bateman oscillators 
is discussed.   The null (or trivial) Lagrangians (NLs), when substituted 
into the Euler-Lagrange (EL) equation, make this equation vanish 
identically, which means that no equation of motion is obtained from
the NLs.  The NLs are also required to be expressed as the total 
derivative of a scalar function [9-13], which is called a gauge 
function [14,15].  

Our objective is to derive the SLs, NLs and their gauge functions (GFs) 
for the Bateman oscillators.  Since these oscillators are non-conservative 
systems, the derived Lagrangians are not consistent with the Helmholtz 
conditions [16,2], which guarantee the existence of Lagrangians for the 
conservative systems.   We present a different view on these conditions 
and their validity for damped (amplified) oscillators.  One obtained SL is 
the CK Lagrangian, which is well-known, and all other NLs and GFs that 
are derived simultaneously with the CK Lagrangian are new.  The CK 
Lagrangian is modified by taking into account the GFs and it is shown 
that this modification allows for the conversion of the undriven Bateman 
oscillators into driven ones, which is our main result.  

The outline of the paper is as follows: the equations of motion for the 
Bateman oscillators, their standard and null Lagrangians, and the gauge 
functions are derived in Section 2; derivation of the modified Caldirola-Kanai 
Lagrangian and the conversion of undriven oscillatory systems and 
driven ones are presented and discussed in Section 3; our conclusions 
are given in Section 4.

\section{Lagrangians and gauge functions for Bateman oscillators}

\subsection{Equations of motion}

The original paper by Bateman [1] considers a non-conservative system
for which the following Lagrangian is proposed 
\[
L_{B} [  \dot x (t), \dot y (t), x (t), y (t)] = m \dot x (t) \dot y (t)
+ {\gamma \over 2} \left [ x (t) \dot y (t) - \dot x (t) y (t) \right ] 
\]
\begin{equation}
\hskip1.7in - k x (t) y (t)\ ,
\label{eq1}
\end{equation}
where $x (t)$ and $y (t)$ are coordinate variables and $\dot x (t)$ and
$\dot y (t)$ are their derivatives with respect to time $t$.  In addition,
$m$ is mass and $\gamma$ and $k$ are damping and spring constants,
respectively.  This Lagrangian is now known as the Bateman Lagrangian 
[2-4] and its substitution into the EL equation for $y(t)$ and $x(t)$ gives 
respectively the resulting equation of motion for the damped oscillator
\begin{equation}
m \ddot x (t) + \gamma \dot x (t) + k x (t) = 0\ ,
\label{eq2}
\end{equation}
and the equation of motion for the amplified oscillator
\begin{equation}
m \ddot y (t) - \gamma \dot y (t) + k y (t) = 0\ .
\label{eq3}
\end{equation}
The equations are uncoupled but they are related to each other by the 
transformation $[x(t), y(t), \gamma] \rightarrow [y(t), x(t), - \gamma]$,
which allows replacing Eq. (\ref{eq2}) by Eq. (\ref{eq3}) and vice versa.

The equations of motion for the damped and amplified oscillators are 
also considered in this paper but as two unrelated dynamical systems 
for which the SLs, NLs and GFs are independently derived; once the 
SLs are obtained for both oscillators, they will be compared to the 
Bateman Lagrangian given by Eq. (\ref{eq1}).     

Let us define $b = \pm \gamma / m$ and $c = k / m = \omega^2_o$, 
where $\omega_o$ is the characteristic frequency of the oscillators, and 
write Eqs (\ref{eq2}) and (\ref{eq3}) as one equation of motion 
\begin{equation}
\ddot x (t) + b \dot x (t) + \omega_o^2 x (t) = 0\ ,
\label{eq4}
\end{equation}
with the understanding that the damped and amplified oscillators require 
$b > 0$ and $b < 0$, respectively, and that the variable $x(t)$ describes
either damped or amplified oscillator.  Let $\hat D = d^2 / dt^2 + b d / dt 
+ \omega_o^2$ be a linear operator, then Eq. (\ref{eq4}) can be written 
in the compact form $\hat D x (t) = 0$.

In the following, we derive the SLs, NLs and GFs for the Bateman oscillators 
whose equations of motion are $\hat D x (t) = 0$.

\subsection{Novel method to derive standard and null Lagrangians}

The first-derivative term in Eq. (\ref{eq4}) can be removed by using 
the standard transformation of the dependent variable [17].  The 
transformation is
\begin{equation}
x (t) =   x_1 (t) e^{- b t / 2}\ ,
\label{eq5}
\end{equation}
where $x_1 (t)$ is the transformed dependent variable, and it gives
\begin{equation}
\ddot x_1 (t) + \left ( \omega_o^2 - {1 \over 4} b^2 \right ) 
 x_1 (t) = 0\ .
\label{eq6}
\end{equation}
Despite the fact that the first derivative term is removed, the coefficient 
$b$ is still present in the transformed equation of motion.  However, if
$b = 0$, then $x_1 (t) = x (t)$ and Eq. (\ref{eq6}) becomes the equation 
of motion for a undamped harmonic oscillator [18,19].

The standard Lagrangian for this equation is 
\begin{equation}
L_{s} [  \dot x_1 (t), x_1 (t)] = {1  \over 2} \left [ \left (\dot x_1 (t) 
\right )^2 - \left ( \omega_o^2 - {1 \over 4} b^2 \right )   x_1^2 (t) \right ]\ ,
\label{eq7a}
\end{equation}
and its substitution into the EL equation gives Eq. (\ref{eq6}).  Let us now follow 
[20,21] and consider the following Lagrangian 
\begin{equation}
L_{n} [  \dot x_1 (t), x_1 (t), t] = C_1 \dot x_1 (t) x_1 (t) + C_2 \left [
\dot x_1 (t) t + x_1 (t) \right ] + C_4 \dot x_1 (t) + C_6\ ,
\label{eq7b}
\end{equation}
where $C_1$, $C_2$, $C_4$ and $C_6$ are constants.  It is easy to verify 
that $L_{n} [  \dot x_1 (t), x_1 (t), t]$ is the null Lagrangian with the
constants being arbitrary, and that this NL is constructed to the {\it lowest 
order of its dynamical variables} [20,21].  The NL can be added to 
$L_{s} [  \dot x_1 (t), x_1 (t)]$ without changing the form of the 
equation of motion (see Eq. \ref{eq6}) resulting from it.

Since the original equations of motion for the Bateman oscillators depend 
on the dynamical variable $x (t)$ not $x_1 (t)$ (see Eqs \ref{eq62} and 
\ref{eq3}), we now use the inverse transform given by Eq. (\ref{eq5})
to convert the variable $x_1 (t)$ into $x (t)$ in both $L_{s} [  \dot x_1 (t), 
x_1 (t)]$ and $L_{n} [  \dot x_1 (t), x_1 (t), t]$.  The resulting Lagrangian is 
\begin{equation}
L [\dot x(t), x(t), t] = L_{CK} [\dot x(t), x(t), t] + L_{n} [\dot x(t), x(t), t]\ ,
\label{eq8}
\end{equation}
where 
\begin{equation}
L_{CK} [\dot x(t), x(t), t] = {1  \over 2} \left [ \left ( \dot x (t) 
\right )^2 - \omega_o^2 x^2 (t) \right ] e^{b t}\ ,
\label{eq9}
\end{equation}
is the Caldirola-Kanai (CK) Lagrangian [7,8], derived here independently;
comparison of the CK and Bateman Lagrangians (Eqs \ref{eq9} and \ref{eq1})
shows significant differences between them.   The presented method gives 
the following null Lagrangian
\begin{equation}
L_{n} [\dot x(t), x(t), t] = \sum_{i = 1}^3 L_{ni} [\dot x(t), x(t), t]\ ,
\label{eq10}
\end{equation}
where the partial null Lagrangians are 
\begin{equation}
L_{n1} [\dot x(t), x(t), t] = \left ( C_1 + {1 \over 2} b \right ) \left [ \dot x (t) 
+ {1 \over 2} b x (t) \right ] x (t) e^{b t}\ ,
\label{eq11}
\end{equation}
\begin{equation}
L_{n2} [\dot x(t), x(t), t] = C_2 \left [ \left ( \dot x (t) + {1 \over 2} b
x (t) \right ) t + x (t) \right ] e^{b t / 2}\ ,
\label{eq12}
\end{equation}
and 
\begin{equation}
L_{n3} [\dot x(t), x(t), t] = C_4 \left [ \dot x (t) + {1 \over 2} b 
x (t) \right ] e^{b t / 2} + C_6\ ,
\label{eq13}
\end{equation}
where $C_1$, $C_2$, $C_4$ and $C_6$ are arbitrary but their physical 
units are different to make all partial null Lagrangians having the same 
units of energy.  These are new null Lagrangians for the Bateman oscillators.  
The fact that $L_{n} [\dot x(t), x(t), t]$ and its partial Lagrangians are the 
NLs can be shown by defining $\hat {EL}$ to be the differential operator 
of the EL equation and verifying that $\hat {EL} (L_{n}) = 0$ as well as 
$\hat {EL} (L_{ni}) = 0$.  It must be also noted that $b = 0$ reduces 
$L_{n} [\dot x(t), x(t), t]$ to the previously obtained null Lagrangian [15].

\subsection{Gauge functions and general null Lagrangians}

For each partial null Lagrangian, its corresponding partial gauge function 
can be obtained and the results are  
\begin{equation}
\phi_{n1} [x(t), t] = {1 \over 2} \left ( C_1 + {1 \over 2} b \right ) x^2 (t) 
e^{b t}\ ,
\label{eq14}
\end{equation}
\begin{equation}
\phi_{n2} [x(t), t] = C_2 x (t) t e^{b t / 2}\ ,
\label{eq15}
\end{equation}
and
\begin{equation}
\phi_{n3} [x(t), t] = C_4 x (t) e^{b t / 2} + C_6 t\ .
\label{eq16}
\end{equation}
These partial gauge functions can be added together to form 
the gauge function 
\begin{equation}
\phi_{n} [x(t), t] = \sum_{i = 1}^3 \phi_{ni} [x(t), t]\ .
\label{eq17}
\end{equation}
The derived gauge function and partial gauge functions reduce
to those previously obtained [15] when $b = 0$ is assumed.

The above partial gauge functions are obtained to the lowest order
of the dynamical variables for the Bateman oscillators.  Therefore, 
the only way to generalize the gauge functions given by Eqs. 
(\ref{eq14}), (\ref{eq15}) and (\ref{eq16}), without changing the
order of their dynamical variables, is to replace the constants $C_1$,
$C_2$, $C_4$ and $C_6$ by the corresponding functions $f_1 (t)$,
$f_2 (t)$, $f_4 (t)$ and $f_6 (t)$, which must be continuous and 
at least twice differentiable and depend only on the independent 
variable $t$.  In order to obey the assumption that our method of
constructing null Lagrangians is limited to the lowest order of the
dynamical variables, the functions cannot depend on the space
coordinates.  An interesting result is that the functions give 
additional degrees of freedom in constructing the null Lagrangians
and their gauge functions.

We follow [20,21] and call these GFs the general GFs and write 
them here as    
\begin{equation}
\phi_{gn1} [x(t), t] = {1 \over 2} \left [ f_1 (t) + {1 \over 2} 
b \right ] x^2 (t) e^{b t}\ ,
\label{eq18}
\end{equation}
\begin{equation}
\phi_{gn2} [x(t), t] = f_2 (t) x (t) t e^{b t / 2}\ ,
\label{eq19}
\end{equation}
and
\begin{equation}
\phi_{gn3} [x(t), t] = f_4 (t) x (t) e^{b t / 2} + f_6 (t) t\ .
\label{eq20}
\end{equation}
The general gauge function is obtained by adding these partial
gauge functions 
\begin{equation}
\phi_{gn} [x(t), t] = \sum_{i = 1}^3 \phi_{gni} [x(t), t]\ .
\label{eq21}
\end{equation}
The results given by Eqs. (\ref{eq18}) through (\ref{eq21})
are new general gauge functions for the Bateman oscillators
and they generalize those previously obtained for Newton's 
law of inertia [20] and for linear undamped oscillators [21]
for which $b=0$.

Using the general gauge function $\phi_{gn} [x(t), t]$, the 
resulting general null Lagrangian can be calculated and the 
result is 
\begin{equation}
L_{gn} [\dot x (t), x(t), t] = \sum_{i = 1}^3 L_{gni} 
[\dot x (t), x(t), t]\ ,
\label{eq22}
\end{equation}
where the partial null Lagrangians are given by 
\[
L_{gn1} [\dot x(t), x(t), t] = \left [ f_1 (t) + 
{1 \over 2} b \right ] \left [ \dot x (t) + {1 \over 2} b x (t) 
\right ] x (t) e^{b t} 
\]
\begin{equation}
\hskip1.25in + {1 \over 2} \dot f_1 (t) x^2 (t) e^{b t}\ ,
\label{eq23}
\end{equation}
\[
L_{gn2} [\dot x(t), x(t), t] = \left [ f_2 (t) \dot x (t) + \dot f_2 (t) 
x (t) \right ] t e^{b t / 2} 
\]
\begin{equation}
\hskip1.25in + \left ( 1 + {1 \over 2} b t \right ) f_2 (t) x (t) e^{b t / 2}\ ,
\label{eq24}
\end{equation}
and 
\[
L_{gn3} [\dot x(t), x(t), t] = \left [ f_4 (t) \dot x (t)  + \dot f_4 (t) x (t) 
\right ]  e^{b t / 2} + {1 \over 2} b f_4 (t) x (t) e^{b t / 2} 
\]
\begin{equation}
\hskip1.25in + \left [ \dot f_6 (t) t + f_6 (t) \right ]\ .
\label{eq25}
\end{equation}

The obtained results show that the replacement of the constants 
$C_1$, $C_2$, $C_4$ and $C_6$ by the functions $f_1 (t)$, $f_2 
(t)$, $f_4 (t)$ and $f_6 (t)$ gives the Lagrangian $L_{gn} [\dot x 
(t), x(t), t]$, which is the null Lagrangian that significantly generalizes 
$L_{n} [\dot x (t), x(t), t]$ given by Eq. (\ref{eq10}).  From a physical 
point of view, it is required that the functions have correct physical units, 
so each partial general Lagrangian has the units of energy.  The functions
may also be further constrained by postulating the invariance of action for
the null Lagrangians and introducing the so-called exact gauge functions
[20], which allows specifying the end points of some of these functions. 

The general null Lagrangian $L_{gn} [\dot x (t), x(t), t]$ and its partial
null Lagrangians reduce to those previously found [20,21] when $b = 0$.
These Lagrangians depend on four functions, which can be constrainted 
by the initial conditions if they are specified.  The initial conditions are 
used to make the action invariant, which requires that the gauge functions
become zero at the initial conditions.  This contrains the values of the 
four functions at the initial conditions, and the gauge functions that make
the action invariant are called the exact gauge functions [20]. 

Since the null Lagrangians do not affect the derivation of the equations of 
motion, their existence is not restricted by the Helmholtz conditions [6,16].  
However, the conditions seem to imply that the Caldirola-Kanai Lagrangian 
cannot be constructed for the Bateman oscillators.  In the following, we 
consider and discuss this problem in detail.

\subsection{Role of the Caldirola-Kanai Lagrangian}

The Caldirola-Kanai Lagrangian, $L_{CK} [\dot x (t), x (t), t]$, when 
substituted into the EL equation, yields $[\hat D x (t)] e^{b t} = 0$, 
which is consitstent with all Helmholtz conditions that are valid for
any system of ordinary differential equations [16].  However, with 
$e^{b t} \neq 0$, the resulting $\hat D x (t) = 0$ does obey the 
first and second Helmholtz conditions but fails to satisfy the third 
condition.  This shows that after the division by $e^{b t}$ the 
equation of motion fails to satisfy the third condition [22]; see 
further discussion below Eq. (\ref{eq29}).

Since the CK Lagrangian depends explicitly on time, the energy 
function [18,19] must be calculated.  Let $E_{CK} [\dot x(t), 
x(t), t]$ be the energy function for $L_{CK} [\dot x (t), x (t), t]$ 
given by
\begin{equation}
E_{CK} [\dot x(t), x(t), t] = \dot x (t) p_c (t) - L_{CK} [\dot x(t), 
x(t), t]\ ,
\label{eq26}
\end{equation}
where the canonical momentum $p_c (t)$ is 
\begin{equation}
p (t) = {{\partial L_{CK} [\dot x(t), x(t), t]} \over {\partial \dot 
x (t)}} = \dot x (t) e^{b t}\ ,
\label{eq27}
\end{equation}
and is different than the linear momentum $p (t) = \dot x (t)$.

Then, the energy function $E_{CK} [\dot x(t), x(t), t]$ can be 
written as  
\begin{equation}
E_{CK} [\dot x(t), x(t), t] = E_{tot} [\dot x (t), x (t)] e^{b t}\ .
\label{eq28}
\end{equation}
where the total energy is 
\begin{equation}
E_{tot} [\dot x (t), x (t)] = {1  \over 2} \left [ \left ( \dot x (t) 
\right )^2 + \omega_o^2 x^2 (t) \right ]\ .
\label{eq29}
\end{equation}
According to Eqs. (\ref{eq28}) and (\ref{eq29}), the energy function
depends explicitly on time through its exponential term, moreover, both
the energy function and total energy do depend on time through $x (t)$.   

Having obtained $E_{CK} [\dot x(t), x(t), t]$, we use 
\begin{equation}
{{d E_{CK} [\dot x(t), x(t), t]} \over {dt}} = - {{\partial L_{CK} 
[\dot x(t), x(t), t]} \over {\partial t}}\ , 
\label{eq30}
\end{equation}
and find the following equation of motion
\begin{equation}
[\hat D x (t)] \dot x (t) e^{bt} = 0\ .
\label{eq31}
\end{equation}
Since $\dot x (t) e^{bt} \neq 0$, Eq. (\ref{eq31}) becomes
$\hat D x (t) = 0$, which is the equation of motion for the 
Bateman oscillators (see Eq. \ref{eq4}) with the well-known
solutions for $x (t)$ [18] that decrease (increase) with time 
as $e^{- bt/2}$ for the damped (amplified) oscillators.

The exponentially decreasing (increasing) in time solutions to
$\hat D x (t) = 0$ make $E_{CK} [\dot x (t), x (t), t]$ = const,
when substituted into Eq. (\ref{eq28}) because of the cancellation 
of $e^{-bt}$ and $e^{bt}$, and show that $E_{tot} [\dot x (t), 
x (t)]$ decreases (increases) exponentially in time for the damped 
(amplified) oscillators.  The fact that $E_{tot} [\dot x (t), x (t)] 
\neq$ const but $E_{CK} [\dot x(t), x(t), t] = E_{tot} [\dot x (t), 
x (t)] e^{b t}$ = const guarantees that the equation of motion 
$[\hat D x (t)] e^{b t} = 0$ resulting from the CK Lagrangian
satisfies the third Helmholtz condition; however, the equation 
$[\hat D x (t)] = 0$ does not satisfy it as is already pointed 
out at the begining of this section. 

The role of the CK Lagrangian in deriving the equation of motion
for the Bateman oscillators has been questioned in the literature 
[23] based on the previous work [24].   The main conclusion of 
that previous research was that the CK Lagrangian does not 
describe the Bateman oscillators but instead a different oscillatory 
system with its mass increasing ($b > 0$) or decreasing ($b < 0$) 
exponentially in time.  However, as pointed out recently in [25],
the previous work has some conceptual errors that led to incorrect 
conclusions.  

The results presented in this paper are consistent with those
given in [25] as we demonstrated that the energy function 
(see Eq. \ref{eq26}) that depends on the canonical momentum 
$p_c (t)$ is constant in time, and that the total energy (see Eq. 
\ref{eq29}) that depends on the linear momentum $p (t)$ is 
not constant in time.  This shows that the total energy and the 
linear momentum decrease (increase) in time for the damped 
(amplified) Bateman oscillators, which is consistent with the 
physical picture of these dynamical systems.  The increase 
(decrease) of the canonical momentum in time does not 
contradict this picture but instead guarantees that the CK 
Lagrangian can be used to derive the equations of motion 
for the Bateman oscillators.

\section{From undriven to driven Bateman oscillators}

\subsection{Total energy function}

Having demonstrated the validity of the CK Lagrangian 
for the Bateman oscillators, we now investigate effects
of the gauge functions on the CK Lagrangian.   Let us 
combine the CK Lagrangian, given by Eq. (\ref{eq9}),
and the general null Lagrangian, given by Eqs. (\ref{eq22}) 
through (\ref{eq25}), together to obtain the following 
Lagrangian $L [\dot x(t), x(t), t] = L_{CK} [\dot x(t), 
x(t), t] + L_{gn} [\dot x(t), x(t), t]$.  By using this 
new Lagrangian, we find the resulting energy function
to be  
\[
E [\dot x(t), x(t), t] = {1  \over 2} \left [ \left ( \dot x (t) 
\right )^2 + \left ( \omega_o^2 - \bar \omega_o^2 (t) \right ) 
x^2 (t) \right ] e^{b t} 
\]
\begin{equation}
\hskip1.25in - F (t) x (t) e^{bt/2} - G (t)\ ,
\label{eq32}
\end{equation}
where
\begin{equation}
\bar \omega_o^2 (t) = \left [ \left ( f_1 (t) + {1 \over 2} b \right ) b + 
\dot f_1 (t) \right ]\ ,
\label{eq33}
\end{equation}
\begin{equation}
F (t) = \left [ \left ( 1 + {1 \over 2} b t \right ) f_2 (t) + 
\dot f_2 (t) t + \dot f_4 (t) + {1 \over 2} b f_4 (t) \right ]\ , 
\label{eq34}
\end{equation}
and 
\begin{equation}
G (t) = \dot f_6 (t) t + f_6 (t)\ .
\label{eq35}
\end{equation}

We make the frequency shift $\bar \omega_o^2$ time-independent
by taking $f_1 (t) = C_1$ and obtaining $\bar \omega_o^2 = (C_1 + 
b / 2) b$.  We also define the total energy
\begin{equation}
E_{tot} [\dot x(t), x(t)] = {1  \over 2} \left [ \left ( \dot x (t) 
\right )^2 + \omega_s^2 x^2 (t) \right ]\ , 
\label{eq36}
\end{equation}
where $\omega_s^2 = \omega_o^2 - \bar \omega_o^2$ = const.  
It must be noted that $E_{tot} [\dot x(t), x(t)]$ is not constant
because $x(t)$ and $\dot x (t)$ decrease exponentially in time if 
$b > 0$ or increase exponentially in time if $b < 0$.   

Using Eq. (\ref{eq36}), we write Eq. (\ref{eq32}) as  
\begin{equation}
E [\dot x(t), x(t), t] = \left [ E_{tot} [\dot x(t), x(t)] -  
F (t) x (t) e^{- bt/2} - G (t) e^{- b t} \right ] e^{b t}\ ,
\label{eq37}
\end{equation}
which shows that the total energy of the system is modified by 
the presence of $F(t)$ and $G(t)$ that depend on the functions 
$f_2 (t)$, $f_4 (t)$ and $f_6 (t)$, which are arbitrary.  Therefore,
these arbitrary functions can be chosen so that the difference 
between $E_{tot} [\dot x(t), x(t)]$ and the two energy terms 
$F (t) x (t) e^{- bt/2}$ and $G (t) e^{- b t}$ approach zero
as $t \rightarrow \infty$.

It must be also noted that in the special case of $b = 0$, Eq. 
(\ref{eq37}) reduces to Eq. (40) derived in [20].  However, if 
$b=0$ and $f_2 (t) = C_2$, $f_4 (t) = C_4$ and $f_6 (t) = 
C_6$, then the total energy function is equal to that obtained 
in [20].

\subsection{Gauge functions and forces}

Despite the fact that the partial null Lagrangians were used in 
deriving the total energy functions, these null Lagrangians
were derived from the partial gauge functions given by Eqs.  
(\ref{eq18}) through (\ref{eq20}).  Based on definitions given 
by Eqs. (\ref{eq33}) through (\ref{eq35}), it can be seen that 
the partial gauge function $\phi_{gn1} [x(t), t]$ contributes 
only to $\bar \omega_o^2 (t)$, and that $F (t)$ is determined 
by $\phi_{gn2} [x(t), t]$ and also by $\phi_{gn3} [x(t), t]$,
which contributes partially to both $F (t)$ and $G (t)$.

Since the energy $E [\dot x(t), x(t), t]$ is modified by the 
presence of the extra energies $F (t) x (t)$ and $G (t)$, a
new Lagrangian corresponding to $E [\dot x(t), x(t), t]$ 
must also be modified, which is achieved by adding these
extra terms to $L_{CK} [\dot x(t), x(t), t]$ given by Eq. 
(\ref{eq9}).  Then, the modified CK Lagrangian $L_{CK,mod} 
[\dot x(t), x(t), t]$ becomes
\[
L_{CK,mod} [\dot x(t), x(t), t] = {1  \over 2} \left [ \left ( 
\dot x (t) \right )^2 - \omega_s^2 x^2 (t) \right ] e^{b t} 
\]
\begin{equation}
\hskip1.25in - F (t) x (t) e^{bt/2} - G (t)\ ,
\label{eq38}
\end{equation}
which is the well-known Lagrangian for forced and damped 
oscillators [18,19].  Our results demonstrate that the gauge 
function is responsible for introducing two extra energy 
terms to the original CK Lagrangian.

Using $L_{CK,mod} [\dot x(t), x(t), t]$, the following equation 
of motion is obtained  
\begin{equation}
[\ddot x (t) + b \dot x (t) +  ( \omega_o^2 - \bar \omega_o^2 
(t)) x (t) ] e^{bt} = F (t) e^{bt/2}\ ,
\label{eq39}
\end{equation}
which can also be written as 
\begin{equation}
\ddot x (t) + b \dot x (t) + \omega_s^2 x (t) = F (t) e^{- bt/2}\ .
\label{eq40}
\end{equation}
An interesting result is that the solutions to the homogeneous 
equation of Eq. (\ref{eq40}) depend on $e^{-bt/2}$, which is 
the exponential factor as in the forcing function; see below for
the full contemplemantary and particular solutions. 

The definition of $F (t)$ shows that it is determined by the gauge 
function $\phi_{gn2} [x(t), t]$, which contributes through its
function $f_2 (t)$ and its derivative, and also partially by 
$\phi_{gn3} [x(t), t]$, which contributes through its function 
$f_4 (t)$ and its derivative.  Moreover, $f_2 (t)$ and $f_4 (t)$ 
can be any functions of $t$ as long as they are differentiable.  
More constraints on these functions can be imposed after 
invariance of the action is considered [20] or the initial 
conditions are specified.

Let us define $\beta = b / 2$ and write Eq. (\ref{eq39}) as
\begin{equation}
\ddot x (t) + 2 \beta \dot x (t) + \omega_s^2 x (t) = F (t) 
e^{- \beta t}\ .
\label{eq41}
\end{equation}
Since $\beta$ = const and $\omega_s$ = const, the solution 
to the homogeneous part of Eq. (\ref{eq40}) is well-known 
[18,19] and given by 
\begin{equation}
x_h (t) = x_o e^{(-\beta + i \omega) t}\ ,
\label{eq42}
\end{equation}
where $x_o$ is an integration constant and $\omega = 
\sqrt{\omega_s^2 - \beta^2}$ is the natural frequency 
of oscillations.  

To find the particular solution $x_p (t)$, the force function
$F(t)$ must be specified.  Let us take 
\begin{equation}
F (t) = F_o e^{(i \Omega) t}\ ,
\label{eq43}
\end{equation}
where $F_o$ and $\Omega$ are the constant amplitude and 
frequency of the driving force, respectively.  Then, we seek 
the particular solution in the following form 
\begin{equation}
x_p (t) = \Gamma e^{(-\beta + i \Omega) t}\ ,
\label{eq44}
\end{equation}
with $\Gamma$ is a constant to be determined.  Substituting 
Eq. (\ref{eq44}) into Eq. (\ref{eq41}) and using Eq. (\ref{eq43}),
we obtain
\begin{equation}
\Gamma = {{F_o} \over {\omega^2 - \Omega^2}}\ .
\label{eq45}
\end{equation}
Thus, the solution to Eq. (\ref{eq40}) is 
\begin{equation}
x (t) = x_o e^{(-\beta + i \omega) t} + {{F_o} \over 
{\omega^2 - \Omega^2}} e^{(-\beta + i \Omega) t}\ .
\label{eq46}
\end{equation}
The main difference between our solution for $x(t)$ and 
that found in textbooks of Classical Mechanics [18,24] is
that both $x_h (t)$ and $x_p (t)$ decay exponentially in 
time, which results in a different $\Gamma$ that shows 
a resonance if $\omega = \Omega$.  

To derive the standard solution given in the textbooks [18,24],
it is required that 
\begin{equation}
F (t) = F_o e^{(\beta + i \Omega) t}\ .
\label{eq47}
\end{equation}
This is allowed as $F(t)$ is arbitrary as long as it is differentiable.  
However, $F(t)$ is expressed in terms of the functions $f_2(t)$ 
and $f_4(t)$ and their derivatives, these two functions are still
to be determined once $F(t)$ is specified.  Since $f_2(t)$ and 
$f_4(t)$ are arbitrary, we may take either $f_2(t) = C_2$ = const 
and solve a first-order ordinary differential equation (ODE) for 
$f_4(t)$ or assume that $f_4(t) = C_4$ = const and solve another 
ODE for $f_2(t)$.  

The important points of the procedure described above are:
(i) Any null Lagrangian has no effect on the equation of motion; 
(ii) Null Lagrangians depend explicitly on time, so when added to 
the standard Lagrangian they make the total Lagrangian to depend 
on time; (iii) The time-dependent Lagrangian requires that the energy 
function is calculated; (iv) The resulting energy function has extra 
energy-like terms that are not null Lagrangians; (v) By adding these 
energy-like terms to the standard Lagrangian, the derived equation 
of motion becomes inhomoheneous (driven).  This shows an interesting 
connection between the null Lagrnagians and their gauge functions, 
and classical forces [20,21].

The presented results demonstrate that the CK Lagrangian modified
by the gauge functions gives the equations of motion for the driven 
Bateman oscillators, which shows that the gauge functions can be 
used to introduce forces to undriven dynamical systems [21].  The
result justifies searches for null Lagrangians and their gauge functions.
Our results concern only the Bateman oscillators but the same method
can be applied to other dynamical systems, whose existing standard 
Lagrangians may give hints how to construct null Lagrangians for these 
systems.  The role of null Lagrangians in classical physics has not yet
been established, with [26,27] being the exceptions.

\section{Conclusions}

We developed the Lagrangian formalism for Bateman oscillators
and used it to derive the Caldirola-Kanai and null Lagrangians. 
For each null Lagrangian the corresponding gauge function was
obtained.  Our results confirm the previous findings [7,8,25] that 
the CK Lagrangian can be used to derive the equations of motion 
for the Bateman oscillators.  We also considered the Helmholtz 
conditions that guarantee the existence of Lagrangians for given 
ODEs, and demonstrated that the energy function obtained from 
the CK Lagrangian is the sufficient condition for the resulting 
equation of motion to satisfy the conditions. 

The results obtained in this paper contradict the previous claims 
[23,24] that the CK Lagrangian is not valid for the Bateman oscillators 
but instead it represents a different dynamical system in which mass 
increases or decreases exponentially in time.  As a result of this 
contradiction, derivations of the Kanai-Caldirola propagator in the 
de Broglie-Bohm theory [28], which are based on the CK Lagrangian 
with its mass being time-dependent [29] must be taken with caution. 

Our main result is that the CK Lagrangian modified by the presence 
of extra energy terms generated by the partial gauge functions can 
be used to describe the driven Bateman oscillators.  The modified 
CK Lagrangian given by Eq. (\ref{eq38}) and the resulting equation 
of motion for the driven Bateman oscillators (see Eq. \ref{eq40}) 
clearly show that the Lagrangian formulation is possible, and that 
the undriven Bateman oscillators can be converted into the driven 
ones if the gauge functions are taken into account.

The Pais-Uhlenbeck [30] and Feshbach-Tikochinsky [31] methods to
quantize the Bateman oscillators are based on the Bateman Lagrangian 
given by Eq. (\ref{eq1}).  Using these methods, it was shown that 
quantization of the damped quantum harmonic oscillator can be done 
in terms of ladder operators [32], which is partially incorrect due to a 
no-go theorem [33].  Moreover, there seems to be also a quantum 
vacuum problem in [32] as demonstrated recently [34].  Quantization 
of the damped harmonic oscillator can also be performed using the 
CK Lagrangian given by Eq. (\ref{eq9}), whose symmetries were 
investigated and constant-of-motion operators that generate the 
Heisenberg-Weyl algebra were introduced [35].  Based of the results 
obtained in this paper, it can also be possible to quantize the driven 
Bateman oscillators by using the modified CK Lagrangian given by 
Eq. (\ref{eq38}), which is suggested for future work.

\bigskip\noindent
{\bf Data Availability}
The data that supports the findings of this study are available 
within the article.

\bigskip\noindent
{\bf Acknowledgments}
We are indebted to three anonymous referees for their valuable 
suggestions that allow us to improve significantly the first version 
of this manuscript.  

\nocite{*}

\end{document}